\documentclass[reprint,amsmath,amssymb,aps,prd,showpacs,twocolumn]{revtex4-1}

\usepackage{epsfig}
\usepackage{dcolumn}
\usepackage{bm}
\usepackage{tikz}
\usepackage{graphicx, graphics, color}
\usepackage{dcolumn}
\usepackage{bm}
\usepackage{hyperref}
\usepackage[mathlines]{lineno}

\newcommand{\be}{\begin{equation}}
\newcommand{\ee}{\end{equation}}
\newcommand{\ben}{\begin{eqnarray}}
\newcommand{\een}{\end{eqnarray}}

\newcommand{\cK}{{\cal K}}

\newcommand{\cO}{{\cal O}}

\newcommand{\cL}{{\cal L}}

\newcommand{\cM}{{\cal M}}

\newcommand{\cU}{{\cal U}}

\newcommand{\p}{\partial}
\newcommand{\na}{\nabla}

\newcommand{\hSi}{{\hat \Sigma}}

\newcommand{\hg}{\hat g}
\newcommand{\hR}{\hat R}

\newcommand{\tg}{\tilde g}

\newcommand{\ep}{\epsilon}

\newcommand{\ga}{\gamma}

\newcommand{\tR}{{\tilde R}}

\newcommand{\tna}{\tilde \na}

\begin{document}

\title{Uniqueness of higher-dimensional Einstein-Maxwell-phantom dilaton field wormholes}
\author{Marek Rogatko} 
\email{rogat@kft.umcs.lublin.pl}
\affiliation{Institute of Physics, 
Maria Curie-Sklodowska University, 
20-031 Lublin, pl.~Marii Curie-Sklodowskiej 1, Poland}

\date{\today}

\begin{abstract}
The uniqueness of static spherically symmetric traversable wormholes with two asymptotically flat ends, subject to the higher-dimensional solutions
of Einstein-Maxwell-phantom dilaton field equations was proved. We considered the case of an arbitrary dilaton coupling constant.
Conformal positive energy theorem plays the key role in the considerations.
\end{abstract}

\pacs{04.20.Jb, 04.40.-b}
\maketitle

\section{Introduction}
Wormholes have been discovered in the realm of general relativity as Einstein-Rosen bridges \cite{ein35}, considered at the beginning of their history 
as geometric models of elementary particles \cite{mis57} or describing the initial data for Einstein equations \cite{mis60}. 
They represent a connection between the two separate parts of the Universe or even join Universes \cite{mor88}, being shortcuts through spacetime.
The simplest models of wormhole, possessing no event horizon and physical singularities, were presented in 
 \cite{ell73}-\cite{ell79}. 
 The requirement for a throat held open is to invoke phantom field, whose energy momentum tensor violates the null, weak and strong energy conditions.
 Moreover the phantom field kinetic energy term is of the reversed sign. 
 Wormholes potential use in space and possibilities of time travels were discussed in \cite{mor88amj,mor88}.
 It turned out that in the generalization of Einstein gravity like Gauss-Bonnet-dilaton theory, wormholes can be built with no use of such exotic kind of matter
 \cite{kan11}-\cite{har13}. Recently, the method of constructing traversable wormholes by applying duality rotation and complex transformations was proposed \cite{gib16,gib17}.
 An electrically charged traversable wormhole solution in Einstein-Maxwell -phantom dilaton gravity, by the assumption that the dilaton is a phantom field, was obtained in \cite{gou18}.
 
 Soon after the stationary axisymmetric wormhole solution were paid attention to \cite{teo98}-\cite{kle14}. It was claimed that the rotating wormholes would be with a higher possibility stable
 \cite{mat06} and therefore traversable.
 
 Due to the possible astrophysical observations of their existence, the system like neutron star wormhole \cite{dzh11,dzh14} or solely wormholes, were considered. The gravitational lensing effects
 caused by these objects were studied in  \cite{cra95}-\cite{jus17}.
 
 Higher-dimensional wormhole solutions were considered especially from the point of view of the unification scheme \cite{bro97}-\cite{tor13}.

On the other hand, one wants to obtain classification of the objects in question, having in mind classification delivered by the black hole uniqueness theorem.
The work in this direction was provided in \cite{rub89},
where the uniqueness theorem or wormhole spaces with vanishing Ricci scalar was derived. In \cite{yaz17} the uniqueness
of Ellis-Bronikov wormhole with phantom field was proposed. The higher-dimensional case of the static spherically symmetric phantom wormholes was treated in \cite{rog18}.
In \cite{laz17} the uniqueness for four-dimensional case of the Einstein-Maxwell -dilaton wormholes with the dilaton coupling constant equal to one, was elaborated.

Our paper will be devoted to higher-dimensional static spherically symmetric wormholes with asymptotically flat ends, being solution of Einstein-Maxwell-phantom
dilaton theory. By the phantom dilaton field we understand the dilaton field with reversed sign in its kinetic term in the underlying action. \\
The studies of black objects solutions in ordinary higher dimensional Einstein-Maxwell-dilaton gravity was motivated by the contemporary unification scheme, M/string theories,
and constituted a highly non-trivial mathematical challenge \cite{gib88}. In string theory, phantom fields appear naturally due to the studies of the so-called {\it negative branes}, e.g., the 
symmetry $SU(N/M)$ can be achieved by stack of $N$-ordinary branes and $M$-{\it negative branes} \cite{vaf01}.

The addition of kinetic term of the gauge or dilaton field with flipping sign leads to unexpected results as
massless black hole  \cite{gib96}. The systematic studies of static black holes in phantom Einstein-Maxwell-dilaton  gravity reveal varieties of asymptotically
flat and non-asymptotically flat solutions with plethora of different types of causal structures \cite{bro06}-\cite{azr11}.
The influence of dark energy (phantom fields) was also widely studied, both analytically and numerically, in the context of gravitational collapse \cite{nak12}-\cite{nak15}. Among all it was shown that in the system
where the dark components are coupled to the electrically charged scalar field via the exponential coupling and the gauge field-Maxwell field kinetic mixing, 
during the collapse in the presence of dark energy dynamical wormholes and naked singularities were formed in emerging spacetimes.

On the other hand, primordial wormholes predicted in the early Universe, where inflation could be driven by dilaton field, might give observational imprints \cite{jus17}. 
In string inspired gravity theories, the exponential coupling to the dilaton field of Gauss-Bonnet term, ensures that this term survives even in four dimensions \cite{kan11,kan12}.

There were also some efforts to study the stability of wormhole solutions in dilatonic generalization of Gauss-Bonnet gravity or Einstein-Maxwell axion theory. It was revealed that a subset of wormhole 
solutions lying close to the border with linearly stable dilatonic black holes in the domain of their existence, is also stable to radial perturbations \cite{kan12}. On the other hand, stability analysis
in dilaton axion Einstein Maxwell gravity concentrated on the analysis of the parameter being coupling function of dilaton field to Maxwell one
\cite{usm10}. It turned out that stability region exists for its value less than one.
Summing it all up, one can see that the theory in question provides interesting and highly non-trivial generalization of Einstein-Maxwell gravity.

 In our considerations we shall elaborate the case
of an arbitrary value of the dilaton coupling constant. The key role in the proof will be played by the conformal positive energy theorem \cite{sim99}.
In order to apply the theorem we use conformal transformations. First one will be implemented to examine the boundary conditions, the other ones to show the nonegativity
of the Ricci scalar. The last one enables us to show the conformal flatness of the spacetime under consideration.

\section{Uniqueness theorem for charged wormholes}
The action describing Einstein-Maxwell-phantom dilaton system may be written as 
\be
S = \int d^n x \sqrt{-g}~ \Big( R + 2 ~\na_\mu \phi \na^\mu \phi  - \ep~e^{- \alpha \phi} F_{\mu \nu} F^{\mu \nu} \Big),
\ee
where $R$ is the Ricci scalar in $n$-dimensional spacetime, while $\na_\alpha$ denotes Levi-Civita connection in the manifold in question.
The constant $\ep = \pm 1$ allows us to introduce in the analysis the phantom $U(1)$-gauge field.
The general $\alpha$-coupling constant is greater than zero and depends on the dimension of the spacetime, i.e.,
$\alpha = \sqrt{8(n-3)/(n-2)}$. 

In what follows we shall  elaborate the static spacetime in the strict sense, with a timelike Killing vector field $\xi_\alpha = (\p/\p t)_\alpha$ determined in each point of the manifold.
The line element of the spacetime in question is provided by
\be
ds^2 = - N^2 dt^2 + g_{ij} dx^i dx^j,
\ee
where by $g_{ij}$ we have denoted the metric tensor of $(n-1)$-dimensional Riemannian manifold. On the other hand, $N$ is a smooth lapse function.
Because of the fact that the lapse function and $g_{ij}$ components are defined on the hypersurface of constant time, they constitute time independent quantities.
One also has that the Maxwell field and phantom scalar are subject to the staticity condition
\be
\cL_{\xi} F_{\mu \nu} = 0, \qquad \cL_\xi \phi = 0,
\ee
where by $\cL_\xi $ the Lie derivative with respect to the Killing vector field $\xi$ is denoted.

The dimensionally reduced equations of motion for Einstein-Maxwell phantom fields are given by
\ben \label{rr}
{}^{(n-1)}R_{ij} &-& \frac{1}{N} {}^{(g)} \na_i {}^{(g)} \na_j N = - 2 {}^{(g)} \na_i  \phi {}^{(g)} \na_j \phi \\ \nonumber \label{np}
&-&2 \ep e^{-\alpha \phi} \Big[ \frac{{}^{(g)} \na_i \psi {}^{(g)} \na^i \psi}{N^2} + g_{ij} \frac{ {}^{(g)} \na_k \psi {}^{(g)} \na^k \psi}{(2-n) N^2} \Big], \\
{}^{(g)} \na_i {}^{(g)} \na^i \phi &+& \frac{{}^{(g)} \na_i N {}^{(g)} \na_i \phi}{N} \\ \nonumber
&+& \frac{\alpha}{2}~ \ep~ e^{-\alpha \phi} \frac{{}^{(g)} \na_k \psi {}^{(g)} \na^k \psi}{N^2} = 0,\\
{}^{(g)} \na_i {}^{(g)} \na^i \psi &-& \frac{{}^{(g)} \na_i N {}^{(g)} \na_i \phi}{N} - \alpha~ {}^{(g)} \na_i \phi {}^{(g)} \na^i \psi =0,\\
{}^{(g)} \na_i {}^{(g)} \na^i N &-& 2\ep~ \frac{(n-3) e^{-\alpha \phi} {}^{(g)} \na_k \psi {}^{(g)} \na^k \psi}{(n-2) N} = 0, 
\een 
where ${}^{(n-1)}R_{ij}$ and ${}^{(g)} \na_i $ are Ricci scalar curvature and connection existing in $(n-1)$-dimensional spacetime. $\psi$ is the electrostatic potential

The strict static spacetime assumption, ensures that we have no event horizons in the manifold in question. Moreover we assume that
the $(n-1)$-dimensional submanifold is complete, i.e., the $(n-1)$-dimensional hypersurfaces of constant time are singularity free.
For a compact subset $\cK \subset {}^{(n-1)}\Sigma$, consisting of two ends ${}^{(n-1)}\Sigma_{\pm}$ diffeomorphic to $R^{n-1}/B^{n-1}$, where $B^{n-1}$
is closed unit ball situated at the origin of $R^{n-1}$, one can find a standard coordinate system in which the
following expansion is proceeded:
\ben \label{a1}
g_{ij} &=& \bigg( 1 + \frac{2}{n-3} \frac{M_\pm}{r^{n-3}} \bigg) \delta_{ij} + \cO \Big(\frac{1}{r^{n-2}}\Big),\\
N &=& N_{\pm} \bigg(1 - \frac{M_\pm}{r^{n-3}} \bigg) +  \cO \Big(\frac{1}{r^{n-2}} \Big),\\
\psi &=& \frac{Q_{\pm}/A}{r^{n-3}} + \cO \Big(\frac{1}{r^{n-2}} \Big),\\ \label{a3}
\phi &=& \phi_{\pm} - \frac{q_\pm}{(n-3) r^{n-3}} + \cO \Big(\frac{1}{r^{n-2}} \Big),
\een
where $N_\pm>0,~\phi_\pm,~\mu_\pm,~Q_\pm,~q_\pm$ are constant.
$M_\pm$ and $q_\pm$ represent the ADM masses and charges $Q_\pm$, up to a constant factor and charges of the two ends ${}^{(n-1)}\Sigma_{\pm}$, while $r^2 = x_m x^m$.
On the other hand, the constant $A$ is given by
\be
A^2 = \frac{2(n-3)}{n-2}.
\ee
The conditions, like strictly static spacetime, completeness of the submanifold, are chosen in such a way that they envisage the properties of the considered wormhole geometry.
The relations (\ref{a1})-(\ref{a3}) constitute the standard notion of asymptotically flat regions.


As was mentioned before, the key role will be played by the conformal positive energy theorem \cite{sim99}.
In order to use it, one ought to have two asymptotically flat 
Riemannian $(n-1)$-dimensional manifolds. let us say, 
$(\Sigma^{(\Phi)},~ {}^{(\Phi)}g_{ij})$ and $(\Sigma^{(\Psi)},~ {}^{(\Psi)}g_{ij})$. 
The metric tensors of the aforementioned two manifolds will be connected by the conformal transformation
of the form ${}^{(\Psi)}g_{ij} = \Omega^2~{}^{(\Phi)}g_{ij}$, where $\Omega$ is a conformal factor.
Then, it yields that the masses of the manifolds in question will satisfy the relation
${}^{(\Phi)}m + \beta~{}^{(\Psi)}m \geq 0$ if ${}^{(\Phi)} R + \beta~\Omega^2~{}^{(\Psi)} R \geq 0$, 
where ${}^{(\Phi)} R $ and ${}^{(\Psi)} R$ are the Ricci scalars with respect to the adequate metric tensors.
$\beta$ is a positive constant. The inequalities are fulfilled if the 
$(n-1)$-dimensional manifolds are flat \cite{sim99}.
One can remark that the conformal positive energy theorem was widely use in proving the uniqueness of higher-dimensional black objects \cite{gib02}-\cite{rog06}.

To commence with, let us consider $(n-1)$-dimensional metric tensor defined by the conformal transformation of the form
\be
{}^{(n-1)}\tg_{ij} = N^{\frac{2}{n-3}} ~g_{ij}.
\ee
The Ricci curvature tensor in the conformally rescaled metric implies
\ben  \label{rij} 
&{}& {}^{(n-1)} \tR(\tg)_{ij} = \frac{1}{N^2} \Big( \frac{n-2}{n-3} \Big) {}^{(n-1)}\tna_i N {}^{(n-1)}\tna_j N \\ \nonumber
&-& 2 {}^{(n-1)}\tna_i \phi {}^{(n-1)}\tna_j \phi - 2 \ep e^{-\alpha \phi} \frac{{}^{(n-1)}\tna_i \psi {}^{(n-1)}\tna_j \psi }{N^2}.
\een
Let us define the quantities as follows:
\ben
\Phi_{\pm 1}&=& \frac{1}{2} \Big[ e^{\frac{\alpha \phi}{2}} N \pm \frac{e^{\frac{\alpha \phi}{2}} }{N} - \frac{A^2 (1 +\ga) e^{-\frac{\alpha \phi}{2}} \ep \psi^2}{N^2} \Big],\\
\Phi_0 &=& A ~(1 + \ga)^{\frac{1}{2}} ~\frac{e^{- \frac{\alpha \phi}{2}} ~\sqrt{\ep}~\psi}{N},\\
\Psi_{\pm 1} &=& \frac{1}{2} \Big( e^{- \frac{ 2 A^2 i \phi}{\alpha}}~N \pm \frac{ e^{\frac{2 A^2 i \phi}{\alpha}}}{N} \Big),
\een
where for the brevity of the notation we set $
 \ga = \alpha^2/{4 A^2}.$
The following symmetric tensors can be defined on the space in question
\ben \nonumber \label{ric}
{}^{(\Phi)} \tR_{ij} &=& {}^{(n-1)}\tna_i \Phi_{-1} {}^{(n-1)}\tna_j \Phi_{-1} - {}^{(n-1)}\tna_i \Phi_0 {}^{(n-1)}\tna_j \Phi_0 \\ 
&-& {}^{(n-1)}\tna_i \Phi_1{}^{(n-1)}\tna_j \Phi_1, \\ \nonumber
{}^{(\Psi)} \tR_{ij} &=& {}^{(n-1)}\tna_i \Psi_{-1} {}^{(n-1)}\tna_j \Psi_{-1} - {}^{(n-1)}\tna_i \Psi_1 {}^{(n-1)}\tna_j \Psi_1.
\een
Defining the metric of the form $\eta_{AB} = diag(1,-1,-1)$, we have that $\Psi_{A} \Psi^{A} = \Phi_{A} \Phi^A = -1$, where $A= (0,~1,-1)$. Consequently, according to the equation (\ref{ric}), one obtains
\ben
{}^{(n-1)}\tna_m {}^{(n-1)}\tna^m \Psi_B  &=& {}^{(\Psi)} \tR_{i}{}{}^{i} ~\Psi_B, \\
{}^{(n-1)}\tna_m {}^{(n-1)}\tna^m \Phi_B  &=& {}^{(\Phi)} \tR_{i}{}{}^{i} ~\Phi_B.
\een
The Ricci curvature tensor of the conformally rescaled metric ${}^{(n-1)}\tg_{ij} $ implies the following relation:
\be
\tR_{ij} = \frac{2}{A^2} (1 + \ga)^{-1}~\Big( {}^{(\Phi)} \tR_{ij}  + \ga ~{}^{(\Psi)} \tR_{ij} \Big).
\ee
Due to the requirement of the conformal positive energy theorem, one introduces
the conformal transformations of the forms as follows:
\be
{}^{(\Phi)}g_{ij}^{\pm} = {}^{(\Phi)}\omega_{\pm}^{\frac{2}{n-3}}~ \tg_{ij},
\qquad
{}^{(\Psi)}g_{ij}^{\pm} = {}^{(\Psi)}\omega_{\pm}^{\frac{2}{n-3}}~ \tg_{ij},
\label{pff}
\ee
where their conformal factors are subject to the relations
\be
{}^{(\Phi)}\omega_{\pm} = \frac{\Phi_{1} \pm 1 }{ 2}, \qquad
{}^{(\Psi)}\omega_{\pm} = \frac{\Psi_{1} \pm 1}{  2}.
\label{pf}
\ee
Having in mind the metric tensors defined by the relation (\ref{pff}), we obtained
four $(n-1)$-dimensional manifolds $(\Sigma^{+ (\Phi)},~ {}^{(\Phi)}g_{ij}^{+})$,
$(\Sigma^{- (\Phi)},~ {}^{(\Phi)}g_{ij}^{-})$, $(\Sigma^{+ (\Psi)},~ {}^{(\Psi)}g_{ij}^{+})$,
$(\Sigma^{- (\Psi)},~ {}^{(\Psi)}g_{ij}^{-})$. Pasting them together \cite{gib02,gib02a} one obtains complete regular hypersurfaces
$\Sigma^{(\Phi)} = \Sigma^{+ (\Phi)} \cup \Sigma^{- (\Phi)}$ and $\Sigma^{(\Psi)} = \Sigma^{+ (\Psi)} \cup \Sigma^{- (\Psi)}$.
${}^{(\Phi)}g_{ij}^{\pm}$ and ${}^{(\Phi)}g_{ij}^{\pm}$ metric are complete. This fact follows from their definitions, completness of $g_{ij}$ and
the inequalities $N_{-}  \le N \le N_+$. The explicit asymptotical behavior of them can be achieved from the asymptotic conditions imposed on $g_{ij}$, lapse function,
electric potential $\psi$ and dilaton field. \\
As in \cite{gib02,gib02a} the resulting manifolds $\Sigma^{(\Phi)} $ and $\Sigma^{(\Psi)} $ are geodesically complete with one asymptotically flat end of vanishing total gravitational mass. 
In what follows we shall denote them respectively by $\Sigma^{(\Phi)}_+ $ and $\Sigma^{(\Psi)}_+$. 

In order to show that the static slice is conformally flat, we implement the conformal positive energy theorem \cite{sim99}.
Therefore we define the other conformal transformation provided by the following relation:
\be
\hg^{\pm}_{ij} = \Big[ \Big( {}^{(\Phi)}\omega_{\pm} \Big)^2
 \Big( {}^{(\Psi)}\omega_{\pm} \bigg)^{2 \ga } \Big]^{\frac{1}{(n-3)(1+ \ga)}}\tg_{ij}.
\ee
It follows that 
the Ricci curvature tensor on the defined space yields
\ben \nonumber \label{ricci}
&{}&(1+ \ga)~\Big[ \Big( {}^{(\Phi)}\omega_{\pm} \bigg)^2
 \bigg( {}^{(\Psi)}\omega_{\pm} \bigg)^{2 \ga } \Big]^{\frac{1}{(n-3)(1+ \ga)}}~\hR^\pm
\\
&=& \Big( {}^{(\Phi)}\omega_{\pm}^{\frac{2}{n-3}} ~{}^{(\Phi)}R^\pm + \ga~
{}^{(\Psi)}\omega_{\pm}^{\frac{2}{n-3}}~ {}^{(\Psi)}R^\pm \bigg) 
\\ \nonumber
&+& \Big( \frac{n-2}{n-3} \Big) \frac{\ga}{\ga +1}
\Big( \tna _{i} \ln {}^{(\Phi)}\omega_{\pm} - \tna _{i} \ln {}^{(\Psi)}\omega_{\pm} \Big)^2. 
\een
It can be found that the first term in the brackets in equation (\ref{ricci}) has the form as
\ben \nonumber \label{ricci1}
&{}& \Big(\frac{n-2}{n-3} \Big)~
{}^{(\Phi)}\omega_{\pm}^{\frac{2}{n-3}} ~{}^{(\Phi)}R^\pm + \ga~
{}^{(\Psi)}\omega_{\pm}^{\frac{2}{n-3}}~ {}^{(\Psi)}R^\pm \\ \nonumber
&=&
\frac{\mid { \Phi_{0} \tna_{i} \Phi_{-1} - \Phi_{-1} \tna_{i} \Phi_{0} } \mid^2}{(\Phi_1 \pm1)^2} \\ 
&+& \ga~
\frac{\mid { \Psi_{1} \tna_{i} \Psi_{-1}
- \Psi_{-1} \tna_{i} \Psi_{1} } \mid^2}{(\Psi_1 \pm 1)^2}.
\een
Just, having in mind the relation (\ref{ricci}) and (\ref{ricci1}), we conclude that $\hR^\pm$ is greater or equal to zero. The application of the conformal positive energy theorem
reveal that 
$(\Sigma^{(\Phi)}_+,~ {}^{(\Phi)}g_{ij})$, $(\Sigma^{(\Psi)}_+,~ {}^{(\Psi)}g_{ij})$ and
$(\hSi_+,~ \hg_{ij})$ are flat and these facts imply that
the conformal factors
${}^{(\Phi)}\omega = {}^{(\Psi)}\omega$ and $\Phi_{1} = \Psi_{1}$, as well as,
$\Phi_{0} = const~ \Phi_{-1}$ and $\Psi_{0} = const~ \Psi_{-1}$. 

All the above provide the conclusion that the manifold $({}^{(n-1)} \Sigma_+,~ g_{ij})$ is conformally flat.
One can rewrite the metric tensor $\hg_{ij}$ in a 
conformally flat form, i.e., we define a function of the form
\be
\hg_{ij} = {\cal U}^{\frac{4}{n-3}}~ {}^{(\Phi)}g_{ij},
\label{gg}
\ee
where one sets ${\cal U} = ({}^{(\Phi)}\omega_{\pm} N)^{-1/2}$.
Because of the fact that the Ricci scalar in $\hg_{ij}$ metric is equal to zero, 
the underlying equations of motion of the system in question reduce
to the Laplace equation on the three-dimensional Euclidean manifold, given by
\be
\na_{i}\na^{i}{\cal U} = 0,
\ee
where $\na$ is the connection on a flat manifold. 

As in scalar phantom uniqueness theorem \cite{rog18}, we can define a local coordinate in the neighborhood $\cM \in \hSi_+$
\be
\delta_{ij}dx^{i}dx^{j} = \rho^{2} d{\cU}^2 + {\tilde h}_{AB}dx^{A}dx^{B},
\ee
where $\rho^2 = \na_b {\cU} \na^b {\cU}.$\\
The manifold in question is totally geodesic, which means that any of its sub-manifold geodesic is a geodesic in the considered manifold.
The other important fact is that the embedding of $\hSi_+$ into Euclidean $(n-1)$-manifold is totally umbilical \cite{kob69}. It results that each connected 
component of $\hSi_+$ is a geometric sphere 
of a certain radius.  Without loss of the generality, the elaborated embedding is also rigid \cite{kob69}, which causes that
we are always able to locate one connected wormhole of a certain radius $\rho$, at $r=r_0$ surface on $\hSi_+$.
The above mathematical construction leads us to a boundary value problem for the Laplace equation on the base space $\Theta = E^{n-1}/B^{n-1}$, with a Dirichlet boundary conditions.
Our system 
is characterized by a parameter which fixes the radius of the inner boundary  and authorizes wormhole of a specific radius $\rho$. 

Let us suppose that we have two solutions being subject to the same boundary value problem. Using the Green
identity and integrating over the volume $\Theta$, one achieves
\ben \nonumber
\Big( \int_{r \rightarrow \infty} - \int_{\Sigma_{wh}} \Big) (\cU_1- \cU_2)~\frac{\p}{\p r} (\cU_1-\cU_2) dS \\
= \int_\Theta \mid \na (\cU_1 - \cU_2) \mid^2 d \Theta.
\een
The left-hand side of the above relation disappears, so we can draw the conclusion that the 
aforementioned two solutions of the Laplace equation with the Dirichlet boundary conditions are identical. It leads us to the main conclusion of our considerations.\\
\noindent
Theorem:\\
Let $\cU_1$ and $\cU_2$ be the two solutions of the Laplace equation on the base space $\Theta = E^{n-1}/B^{n-1}$, as defined above.
They constitute solution of Einstein-Maxwell-phanton dilaton equations of motion, describing static spherically symmetric traversable wormholes  with two asymptotically flat ends.
The solutions in question are subject to the same boundary and regularity conditions. Then, $\cU_1 = \cU_2$ in all of the region of the base space $\theta$, provided that
$\cU_1(p) = \cU_2(p)$ for at least one point belonging to the aforementioned region.\\

\section{Conclusions}
We considered $n$-dimensional static spherically symmetric wormhole with two asymptotically flat ends, being the solution of Einstein-Maxwell-phantom dilaton theory,
with arbitrary dilaton coupling constant. In our analysis we keep  the constant $\ep = \pm 1$, in front of the kinetic term of gauge field, which allows us to introduce into the 
investigations also the phantom $U(1)$-gauge field.  Our main tools in the proof were conformal transformations and the conformal positive energy theorem \cite{sim99}.

It will be also interesting to pay attention to higher-dimensional rotating wormhole solution with or without $U(1)$-gauge fields. We hope to return to these problems elsewhere.

\begin{acknowledgments}
 {MR was partially supported by the grant DEC-2014/15/B/ST2/00089 of the National Science Center.}
 \end{acknowledgments}



\end{document}